\documentclass[UTF8]{iopart}
\usepackage{graphicx}% Include figure files

\begin{document}

\title{Universal dynamics of zero-momentum to plane-wave transition in spin-orbit coupled Bose-Einstein condensates}

\author{Qinzhou Ye, Shuyuan Wu, Xunda Jiang and Chaohong Lee}

\address{$^1$ Laboratory of Quantum Engineering and Quantum Metrology, School of Physics and Astronomy, Sun Yat-Sen University (Zhuhai Campus), Zhuhai 519082, China}
\address{$^2$ Key Laboratory of Optoelectronic Materials and Technologies, Sun Yat-Sen University (Guangzhou Campus), Guangzhou 510275, China}
\ead{lichaoh2@mail.sysu.edu.cn}
\vspace{10pt}

\begin{abstract}
 We investigate the universal spatiotemporal dynamics in spin-orbit coupled Bose-Einstein condensates which are driven from the zero-momentum phase to the plane-wave phase.
 The excitation spectrum reveals that, at the critical point, the Landau critical velocity vanishes and the correlation length diverges.
 Therefore, according to the Kibble-Zurek mechanism, spatial domains will spontaneously appear in such a quench through the critical point.
 By simulating the real-time dynamics, we numerically extract the static correlation length critical exponent $v$ and the dynamic critical exponent $z$ from the scalings of the temporal bifurcation delay and the spatial domain number.
 The numerical scalings consist well with the analytical ones obtained by analyzing the excitation spectrum.
\end{abstract}

\vspace{2pc}
\noindent{\it Keywords}: Bose-Einstein condensation, quantum criticality, quantum phase transitions, quantum quenches
%\maketitle

\section{Introduction}

The critical behavior near a continuous phase transition has been explored in many areas of physics, including cosmology, particle physics and condensed matter.
When a system is driven across a continuous phase transition point, both the relaxation time and correlation length diverge at the critical point, so that the time-evolution cannot be adiabatic no matter how slow the quench is.
Therefore, for a quench with finite quench rate, the system will go out of equilibrium near the critical point and defects spontaneously form.
The Kibble-Zurek mechanism (KZM)~\cite{Kibble_JPMG_9_1387,Zurek_Nature_317_505,Dziarmaga_AP_59_1063,Campo_IJMPA_29_1430018} provides a general theory for understanding the non-equilibrium dynamics crossing the critical point and predicts universal scaling laws of the defect density with respect to the quench rate.
%The KZM scaling has been confirmed in both classical phase transitions and quantum phase transitions (QPTs). [chapman][Dziamaga Advances in physics][clark]
The KZM have been found in various systems, such as the early universe~\cite{Kibble_JPMG_9_1387}, superfluid helium~\cite{Zurek_Nature_317_505}, liquid crystal\cite{Nikkhou_NP_11_183} and ion crystal\cite{Ulm_NC_4_2290,Pyka_NC_4_2291}.
Recently, due to the extraordinary degree of flexibility and high controllability, atomic Bose-Einstein condensates (BECs) becomes an excellent candidates for exploring the KZM, in both thermodynamic~\cite{Damski_PRL_104_160404,Das_SR_2_352,Donner_Science_315_1556,
Lamporesi_NP_9_656,Navon_Science_347_167,SuS_PRL_110_215302,Weiler_Nature_455_948,
Witkowska_PRL_106_135301} and quantum phase transitions~\cite{Uhlmann_PRL_99_120407,Saito_PRA_76_043613,Damski_PRL_99_130402,
Dziarmaga_PRL_101_115701,LeeC_PRL_102_070401,ChenD_PRL_106_235304,
Sabbatini_PRL_107_230402,Anquez_PRL_116_155301,Clark_Science_354_606,
XuJ_EPL_113_50003,WuS_PRA_94_043606,KangS_PRA_95_053638,WuS_PRA_95_063606}.

%###############
In recent years, one remarkable advance in cold atom research is the realization of spin-orbit (SO) coupling.
In the pioneering experiments, the SO coupling is created with two spin states of $^{87}Rb$ coupled by two counter propagating Raman lasers~\cite{LinY_Nature_471_83}.
Due to the competition of the SO coupling and the atom-atom interactions, various novel superfluid phases emerge.
The ground-state phase diagram is predicted to include a stripe phase, a plane-wave phase and a zero-momentum phase~\cite{HoT_PRL_107_150403,LiY_PRL_108_225301}.
A great amount of experimental and theoretical efforts have been dedicated to study the ground-state properties and static phase transitions~\cite{HuH_PRL_108_010402,Ozawa_PRL_109_025301,Galitski_Nature_494_49,
JiS_NP_10_314,Hamner_NC_5_4023,ZhaiH_RPP_78_026001}.
The phase transition dynamics and the critical scaling behavior, however, have rarely been explored.
In fact, the high tunability of Raman coupling parameters makes SO coupled BECs an ideal platform to test the KZM.
Taking advantage of the simplicity of experimental setup and the developed techniques, we suppose that SO coupled two-component BECs could be a good choice to explore the critical behavior.

In this paper, we investigate the KZM in SO coupled two-component BECs by both analyzing the Bogoliubov excitations and the population dynamics during the phase transition.
We find spatial domains form spontaneously after quenching through critical point due to the vanish of Landau critical velocity and the divergence of correlation length.
The KZ scalings can be extracted from the excitation spectrum and the scaling of Landau critical velocity.
On the other hand, by simulating the real-time dynamics, we find that the average bifurcation delay and average domain number follow universal scaling laws given by the KZM.
The critical exponents derived from the numerical results consist well with the analytical ones.

The paper is organized as follows. In Sec. \uppercase\expandafter{\romannumeral2}, we describe the model and ground-state phase diagram. We also give a brief introduction to the KZM.
In Sec. \uppercase\expandafter{\romannumeral3}, we calculate the Bogoliubov excitations to study the Landau critical velocity and the critical scalings.
In Sec. \uppercase\expandafter{\romannumeral4}, we show how to extract the critical exponents from the real-time dynamics.
In Sec. \uppercase\expandafter{\romannumeral5}, we summarize our results and briefly discuss the experimental feasibility.

\section{Zero-momentum to plane-wave transition and Kibble-Zurek mechanism}

We consider a pseudo-spin-1/2 atomic Bose gas with Raman process induced SO coupling along x-direction~\cite{LinY_Nature_471_83}. For simplicity, we restrict our discussion to an elongated system where the dynamics are confined to the x-direction and assume that the system remains in the ground-states in the transverse directions.
The single particle effective Hamiltonian in the pseudo-spin-1/2 basis $\Psi=(\psi_1, \psi_2)^T$ can be written as (set $\hbar=m=1$):
\begin{eqnarray}
\label{single_Hamiltonian}
h_0=\frac{1}{2}(k_x-k_r\sigma_z)^2+\frac{\Omega}{2}\sigma_x
+\frac{\delta}{2}\sigma_z.
\end{eqnarray}
Here, $k_r$ is the recoil momentum of Raman coupling, $\Omega$ is the Raman coupling strength, $\delta$ is the Raman detuning, and $\sigma_{x,y,z}$ are $2\times2$ Pauli matrices. In the following, energy is conveniently measured in units of the recoil energy $E_r=k_r^2/2$.

Considering also the atom-atom interactions, the mean-field (MF) energy functional of the system can be expressed as:
\begin{eqnarray}
\label{energy_func}
&E=E_0+E_I,\nonumber \\
&E_0=\int\,dx
\left(
  \begin{array}{cc}
    \psi_1^\ast & \psi_2^\ast
  \end{array}
\right)h_0
\left(
  \begin{array}{c}
    \psi_1 \\ \psi_2
  \end{array}
\right),\nonumber\\
&E_I=\int\,dx
\frac{g_{11}}{2}\left|\psi_1\right|^4+\frac{g_{22}}{2}\left|\psi_2\right|^4
+g_{12}\left|\psi_1\right|^2\left|\psi_2\right|^2.
\end{eqnarray}
Here, $g_{11}$ and $g_{22}$ are the intra-species interaction strength and $g_{12}$ is the inter-species interaction strength, which are determined by the intra-species and inter-species $s$-wave scattering lengths respectively.
In the following parts, we consider only the spin-symmetry interaction $(g_{11}=g_{22}=g)$ and the resonance case $(\delta=0)$ for simplification.
Therefore the interaction energy can be rewritten as:
\begin{equation}
\label{interaction_energy}
E_I=\int\,dx\left[
\frac{g}{2}\left(
n_1+n_2
\right)^2+(g_{12}-g)n_1n_2
\right]
\end{equation}
in which $n_1=\left|\psi_1\right|^2$ and $n_2=\left|\psi_2\right|^2$ are the densities for the two spin components.

To obtain the ground-state wave-function $\phi=(\phi_1,\phi_2)^T$, one can utilize the variational method and adopt the variational ansatz~\cite{LiY_PRL_108_225301}:
\begin{eqnarray}
\label{groundstate_ansatz}
\left(
  \begin{array}{c}
    \!\phi_1 \\ \!\phi_2
  \end{array}
\right)\!=
\sqrt{\bar{n}}
\left[
C_a\!\left(
  \begin{array}{c}
    \!\cos\theta \\ \!-\sin\theta
  \end{array}
\right)e^{ik_1x}\!+
C_b\!\left(
  \begin{array}{c}
    \!\sin\theta \\ \!-\cos\theta
  \end{array}
\right)e^{-ik_1x}
\right],
\end{eqnarray}
in which $\bar{n}=N/L$ is the average atom density with $N$ and $L$ being the total atom number and the size of the system in x-direction, and $C_{a,b}$, $\theta$ and $k_1$ are the variational parameters. The normalization condition indicates $\left|C_a\right|^2+\left|C_b\right|^2=1$.
By inserting the condensates wave-function~(\ref{groundstate_ansatz}) into the energy functional~(\ref{energy_func}) and minimizing the energy with respect to the variational parameters, one can obtain the ground-state for given $\Omega$ and interaction strength $g$ and $g_{12}$.
The ground-state phase diagram has been discussed in the work by Li \emph{et al.}~\cite{LiY_PRL_108_225301} and we just give a brief summary here.

(\uppercase\expandafter{\romannumeral1}) When $\Omega$ is relatively small, the energy functional~(\ref{energy_func}) has two degenerate minima at $k=\pm k_1$ and the condensate wave-function is the superposition of these two quasimomentum components, namely $C_a\neq0$ and $C_b\neq0$ in the ansatz~(\ref{groundstate_ansatz}).
Therefore the densities of both spin components have spatial modulation.
This is named as a stripe condensate.

(\uppercase\expandafter{\romannumeral2}) As $\Omega$ increases, the density modulation in the stripe phase increases and the density-density term $\frac{g}{2}(n_1+n_2)^2$ in the interaction energy~(\ref{interaction_energy}) cost more and more energy. When $\Omega$ exceeds a critical value $\Omega_1$, the minimization of energy functional~(\ref{energy_func}) gives either $C_a=0$ or $C_b=0$.
The condensate wave-function has a single quasimomentum component, which is named as a plane-wave (PW) condensate.

(\uppercase\expandafter{\romannumeral3}) If $\Omega$ further increases and exceeds another critical value $\Omega_C$,
the two minima at $k=\pm k_1$ will emerge into a single minimum at $k=0$. Then the condensate wave-function is a plane-wave with zero quasimomentum (ZM), which is named as a ZM condensate.

(\uppercase\expandafter{\romannumeral4}) If the average atom density $\bar{n}$ exceeds a critical value $n_c$, the stripe condensate always have a lower energy than the PW condensate. Therefore, there will be a direct transition from the stripe phase to the ZM phase when $\bar{n}>n_c$.

Here we concentrate on the transition from the ZM phase to the PW phase, which is a second-order phase transition since a single minimum in the energy dispersion splits into two minima with the quasimomentum changing continuously.
The critical coupling strength for the transition from the ZM phase to the PW phase is
\begin{equation}
\label{critical_coupling}
\Omega_C=2(k_r^2-2G),
\end{equation}
in which $G=\bar{n}(g-g_{12})/4$;
in the PW phase the two minima locate at
\begin{equation}
k=\pm k_1=\pm k_r\sqrt{1-\frac{\Omega^2}{4(k_r^2-2G)^2}};
\end{equation}
and the variational parameter $\theta$ in the ground-state wave-function~(\ref{groundstate_ansatz}) is
\begin{equation}
\theta=\arccos(k_1/k_r)/2.
\end{equation}
In addition, to characterize different phases, we can define the spin polarization as:
\begin{equation}
\label{porlarization}
\mathcal{P}=\int\,dx J_z(x),
\end{equation}
in which $J_z(x)=[n_1(x)-n_2(x)]/[n_1(x)+n_2(x)]$.
The ZM condensate has a zero $\mathcal{P}$ while PW condensate has a nonzero $\mathcal{P}$.

In the SO coupled BECs, by adjusting the coupling strength $\Omega$ from an initial value $\Omega_i>\Omega_C$ to a final value $\Omega_f<\Omega_C$, the system is driven across the critical point of the continuous phase transition; see the inset of figure~\ref{excitation_spectrum}.
The corresponding nonequilibrium dynamics are expected to show a universal scaling behavior according to the KZM.
It is convenient to define the dimensionless distance from the critical point as
\begin{equation}
\label{dimensionless_distance}
\epsilon(t)=\frac{\Omega(t)-\Omega_C}{\Omega_C}.
\end{equation}
We linearly quench the Raman coupling strength $\Omega$, so that the dimensionless distance varies linearly near the critical point as
\begin{equation}
\epsilon(t)=-\frac{t}{\tau_Q},
\end{equation}
in which $\tau_Q$ is the quench time.
When $\epsilon(t)\to0$, the correlation length $\xi$ and the relaxation time $\tau_r$  of the system diverge as~\cite{Dziarmaga_AP_59_1063}:
\begin{equation}
\xi\propto\left|\epsilon\right|^{-\nu}, \tau_r\propto\xi^z\propto\left|\epsilon\right|^{-\nu z},
\end{equation}
where $z$ and $\nu$ are the critical exponents.

According to the KZM, the evolution of the system during the quench can be divided into three stages decided by two characteristic time scales. One time scale is the relaxation time $\tau_r$, which characterizes how fast the system follows the ground-state of its instantaneous Hamiltonian. The other time scale is the transition time $\tau_t=\epsilon(t)/\dot{\epsilon}(t)$, which describes how fast the time-dependent parameter changes.
Initially, when the system is far away from the critical point, the relaxation time $\tau_r$ is shorter than the transition time $\tau_t$ so that the system can follow the instantaneous ground-state adiabaticlly.
Because of the divergence of $\tau_r$ near the critical point, the transition from the adiabatic stage to the impulse stage happens when the two time scales become comparable, which defines the freezing time $-\hat{t}$ according to
\begin{equation}
\tau_r(-\hat{t})=\epsilon(-\hat{t})/\dot{\epsilon}(-\hat{t}).
\end{equation}
In the impulse stage, the system becomes effectively frozen and stays in the instantaneous ground-state of the time $-\hat{t}$.
When the two time scales become comparable again at $\hat{t}$, which is called freeze-out time, the adiabatic evolution of the state restarts.
At the freeze-out time, the dimensionless distance $\hat{\epsilon}$ and the correlation length $\hat{\xi}$ both have power-law scalings as a function of the quench time $\tau_Q$:
\begin{equation}
\label{scaling_1}
\hat{\epsilon}=\epsilon(\hat{t})\propto\tau_Q^{-\frac{1}{1+\nu z}}, \hat{\xi}=\xi(\hat{t})\propto\tau_Q^{\frac{v}{1+\nu z}}.
\end{equation}
After the impulse-adiabatic transition, the system locates in the PW phase, which has two degenerate ground-states with different nonzero quasimomenta.
Therefore the system choose the state randomly in the space and domains appear.
Since the domains at a distance larger than the correlation length $\hat{\xi}$ form independently, the average domain number at $\hat{t}$ scales as:
\begin{equation}
\label{scaling_2}
N_d(\hat{t})\propto \hat{\xi}^{-d}\propto\tau_Q^{-\frac{d\nu}{1+\nu z}},
\end{equation}
where $d$ is the number of space dimensions.

In the following, we will explore the universal critical dynamics and derive the critical exponents $z$ and $\nu$ through two complementary approaches, by analyzing the Bogoliubov excitation spectrum and by performing numerical simulation of the real-time dynamics.

\section{Spontaneous superfluidity breakdown near the critical point}

In this section, we investigate the universal scaling by analyzing the spontaneous breakdown of superfluidity.
According to the Landau criterion, if the superfluid velocity is less than the Landau critical velocity, the elementary excitations is prohibited due to the conservation of energy and momentum.
However, around a continuous phase transition, the Landau critical velocity vanishes and elementary excitations appear spontaneously.
We will show that the scaling exponents can be extracted from the excitation spectrum and the Landau critical velocity.

\begin{figure}[htb]
\begin{center}
\includegraphics[width=10 cm]{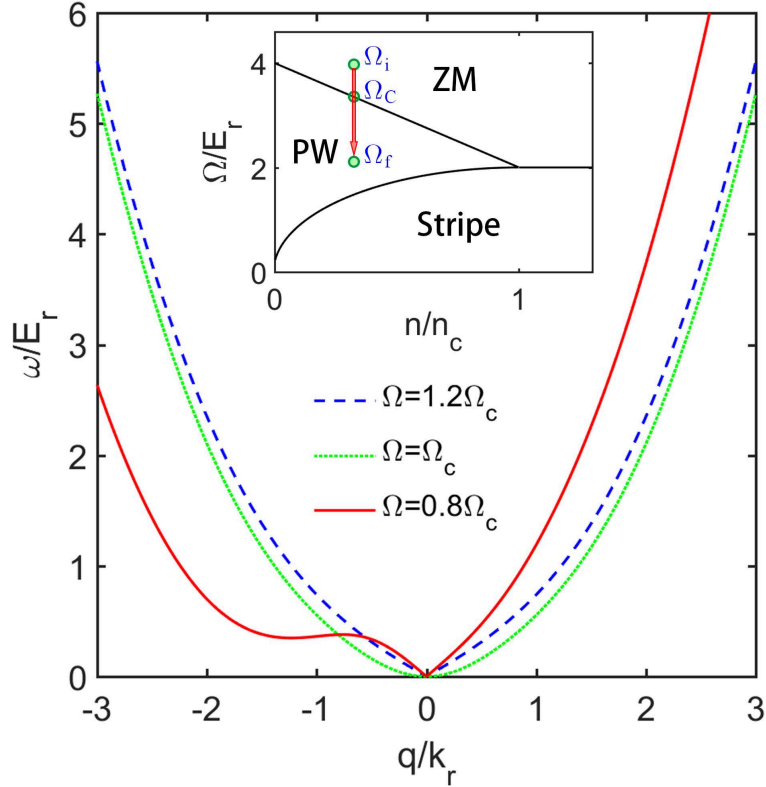}
\caption{\label{excitation_spectrum} (Color online) Typical excitation spectra for system in the ZM phase (dashed line), the critical point (dotted line) and the PW phase (solid line). The inset shows the ground-state phase diagram and the quenching of Raman coupling strength $\Omega$, where $\Omega_C$ is the critical point and $\Omega_i$ ($\Omega_f$) stands for the initial (final) Raman coupling strength. The parameters: $gn/E_r=1$, $\gamma=(g-g_{12})/(g+g_{12})=0.0012$.}
\end{center}
\end{figure}

Firstly, we perform a Bogoliubov analysis to obtain the excitation modes over the MF ground-states. By minimizing the MF energy functional~(\ref{energy_func}) with respect to the wave-functions $\psi_{1,2}$, one obtains two-component time-dependent Gross-Pitaevskii equations (GPEs), which describe the dynamics of the system.  The GPEs read as:
\begin{eqnarray}
\label{GPEs}
i\frac{\partial\psi_1}{\partial t}=\left(-\frac{1}{2}\partial_x^2+ik_r\partial_x+\frac{k_r^2}{2}\right)\psi_1
+\frac{\Omega}{2}\psi_2+g\left|\psi_1\right|^2\psi_1
+g_{12}\left|\psi_2\right|^2\psi_1, \nonumber\\
i\frac{\partial\psi_2}{\partial t}=\left(-\frac{1}{2}\partial_x^2-ik_r\partial_x+\frac{k_r^2}{2}\right)\psi_2
+\frac{\Omega}{2}\psi_1+g_{12}\left|\psi_1\right|^2\psi_2
+g\left|\psi_2\right|^2\psi_2.\nonumber\\
\end{eqnarray}
In the PW and ZM phase, the condensate wave-function can be expanded as:
\begin{equation}
\left(
  \begin{array}{c}
    \!\psi_1 \\ \!\psi_2
  \end{array}
\right)\!=
e^{-i\mu t}\left(
  \begin{array}{c}
    \!\phi_1 \\ \!\phi_2
  \end{array}
\right),
\left(
  \begin{array}{c}
    \!\phi_1 \\ \!\phi_2
  \end{array}
\right)\!=
\sqrt{\bar{n}}\left(
  \begin{array}{c}
    \!C_1 \\ \!C_2
  \end{array}
\right)e^{ik_1x},
\end{equation}
in which $\mu$ is the chemical potential and $C_{1,2}$ are the wave-function amplitudes.
In the PW phase, the ground-state with quasimomentum $k_1$ has wave-function amplitudes $(C_1=\cos\theta, C_2=-\sin\theta)$ while the ground-state with quasimomentum $-k_1$ has wave-function amplitudes $(C_1=\sin\theta, C_2=-\cos\theta)$.
For simplicity, we choose the ground-state with the quasimomentum $k_1$ in calculating the excitation spectrum for the PW phase.
In the ZM phase, the ground-state has $(C_1=\sqrt{2}/2,C_2=-\sqrt{2}/2)$.
To determine the Bogoliubov excitation spectrum, we consider small perturbations around the ground-state
\begin{equation}
\label{perturbed_groundstate}
\left(
  \begin{array}{c}
    \!\psi_1 \\ \!\psi_2
  \end{array}
\right)\!=
e^{-i\mu t}\left[
\left(
  \begin{array}{c}
    \!\phi_1 \\ \!\phi_2
  \end{array}
\right)+
\left(
  \begin{array}{c}
    \!\delta\phi_1(x,t) \\ \!\delta\phi_2(x,t)
  \end{array}
\right)
\right].
\end{equation}
Inserting equation~(\ref{perturbed_groundstate}) into the equations~(\ref{GPEs}), one obtains the linearized equations for the perturbations:
\begin{eqnarray}
\label{linearized_Eq}
i\frac{\partial}{\partial t}\delta\phi_1&=\left(-\frac{1}{2}\partial_x^2
+ik_r\partial_x+\frac{k_r^2}{2}-\mu\right)\delta\phi_1
+\frac{\Omega}{2}\delta\phi_2
+g\left(2\left|\phi_1\right|^2\delta\phi_1+\phi_1^2\delta\phi_1^\ast\right)\nonumber
\\&+g_{12}\left(\left|\phi_2\right|^2\delta\phi_1+\phi_1\phi_2^\ast\delta\phi_2+\phi_1\phi_2\delta\phi_2^\ast\right), \\
i\frac{\partial}{\partial t}\delta\phi_2&=\left(-\frac{1}{2}\partial_x^2
-ik_r\partial_x+\frac{k_r^2}{2}-\mu\right)\delta\phi_2
+\frac{\Omega}{2}\delta\phi_1
+g\left(2\left|\phi_2\right|^2\delta\phi_2+\phi_2^2\delta\phi_2^\ast\right)\nonumber
\\&+g_{12}\left(\left|\phi_1\right|^2\delta\phi_2+\phi_1^\ast\phi_2\delta\phi_1+\phi_1\phi_2\delta\phi_1^\ast\right).
\end{eqnarray}
The perturbations $\delta\phi_{1,2}$ can be written in the form
\begin{equation}
\label{perturbation}
\left(
  \begin{array}{c}
    \!\delta\phi_1 \\ \!\delta\phi_2
  \end{array}
\right)\!=\!
\left(
  \begin{array}{c}
    \!u_{1,q} \\ \!u_{2,q}
  \end{array}
\right)\!e^{ik_1x+iqx-i\omega t}\!+\!
\left(
  \begin{array}{c}
    \!v_{1,q}^\ast \\ \!v_{2,q}^\ast
  \end{array}
\right)\!e^{ik_1x-iqx+i\omega t},
\end{equation}
in which $q$ is the excitation momentum, $\omega$ is the excitation frequency, and $u_{j,q}$ and $v_{j,q}, j=1,2$ are the complex amplitudes. Substituting equation~(\ref{perturbation}) to the linearized equations~(\ref{linearized_Eq}) and comparing the coefficients for the terms of $e^{iqx-i\omega t}$ and $e^{-iqx+i\omega t}$, one can obtain the Bogoliubov-de-Gennes (BdG) equations:
\begin{equation}
\label{BdG}
\mathcal{M}(q)
\left(
  \begin{array}{c}
    \!u_q \\ \!v_q
  \end{array}
\right)=\omega
\left(
  \begin{array}{c}
    \!u_q \\ \!v_q
  \end{array}
\right),
\end{equation}
in which $u_q=\left(u_{1,q}, u_{2,q}\right)^T$, $v_q=\left(v_{1,q}, v_{2,q}\right)^T$ and
\begin{equation}
\mathcal{M}(q)=
\left(
  \begin{array}{cc}
    A_+-\mu+B & C \\ -C^\ast & -A_-+\mu-B^\ast
  \end{array}
\right)
\end{equation}
with
\begin{eqnarray}
&A_\pm=
\left[
  \begin{array}{cc}
   (k_1\pm q-k_r)^2/2 & 0 \\ 0 & (k_1\pm q+k_r)^2/2
  \end{array}
\right],\nonumber\\
&B=
\left[
  \begin{array}{cc}
    2g\bar{n}\left|C_1\right|^2+g_{12}\bar{n}\left|C_2\right|^2 & \Omega/2+g_{12}\bar{n}C_1C_2^\ast \\ \Omega/2+g_{12}\bar{n}C_1^\ast C_2 & g_{12}\bar{n}\left|C_1\right|^2+2g\bar{n}\left|C_2\right|^2
  \end{array}
\right],\nonumber\\
&C=
\left[
  \begin{array}{cc}
    g\bar{n}C_1^2 & g_{12}\bar{n}C_1C_2 \\ g_{12}\bar{n}C_1C_2 & g\bar{n}C_2^2
  \end{array}
\right].
\end{eqnarray}
Then the excitation spectrum can be obtained by diagonalizing the matrix $\mathcal{M}(q)$. Three typical excitation spectra for the system in the PW phase, the ZM phase and the critical point are shown in figure~\ref{excitation_spectrum}.

From figure~\ref{excitation_spectrum}, we see that the excitation spectra exhibit phonon modes with linear dispersions in long wavelength limit for both PW phase and ZM phase, namely $\omega(q)=-c_1q$ for $q<0$, and $\omega(q)=c_2q$ for $q>0$, where $c_1$ ($c_2$) is the sound velocity in the negative (positive) x-direction.
The phonon modes in the excitation spectrum are significant features of superfluidity.
Interestingly, at the critical point between the PW and ZM phases, softening of the phonon modes is observed and the elementary excitations exhibit a $q^2$ dependence (see the dotted line in figure~\ref{excitation_spectrum}), which is due to the divergency of the effective mass associated with the single particle spectrum at the critical point~\cite{JiS_PRL_114_105301}.
Since $\omega(q)\propto\left|q\right|^z$ as $q\to 0$ at a continuous phase transition~\cite{Sachdev_book,Robinson_book,Polkovnikov_RMP_83_863}, we have the dynamical critical exponent $z=2$.

\begin{figure}[htb]
\begin{center}
\includegraphics[width=\columnwidth]{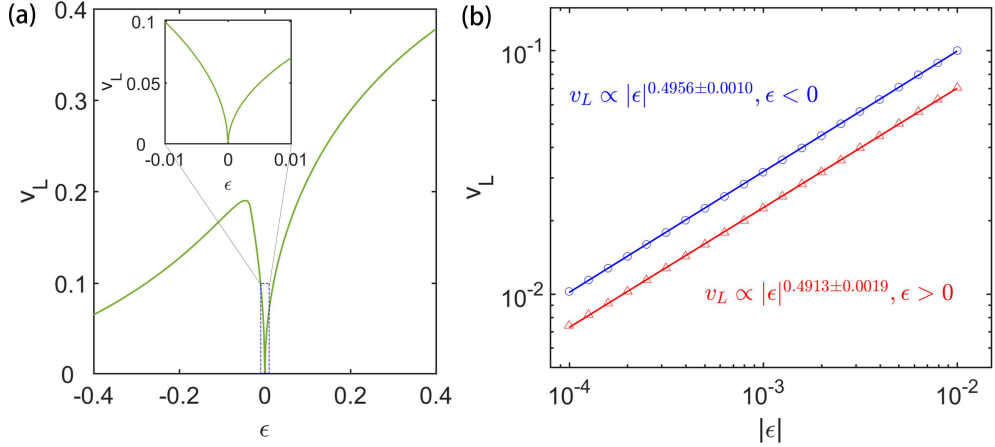}
\caption{\label{Scaling_vl} (Color online) (a) The Landau critical velocity for different $\epsilon=(\Omega-\Omega_c)/\Omega_c$. The inset shows a zoom of $v_L$ near the critical point. (b) The scaling of $v_L$ versus $\left|\epsilon\right|$ near the critical point. The blue circles and red triangles represent the Landau critical velocity $v_L$ in the PW and ZM phases respectively. The solid lines are the linear fittings. The parameters: $gn/E_r=1$, $\gamma=(g-g_{12})/(g+g_{12})=0.0012$. }
\end{center}
\end{figure}

The Landau critical velocity
\begin{equation}
v_L=\mathop{\min}_q\left|\omega/q\right|
\end{equation}
for different $\epsilon$ can be directly extracted from the excitation spectrum.
As illustrated in figure~\ref{Scaling_vl}a, the Landau critical velocity $v_L$ vanishes at the critical point $\epsilon=0$, which is due to the softening of the phonon modes, namely $\omega(q)\propto\left|q\right|^2$ as $q\to 0$.
Therefore, elementary excitations can appear spontaneously around the phase transition.
In the ZM phase, the Landau critical velocity $v_L$ equals with the sound velocity and increase monotonously with increasing $\left|\epsilon\right|$.
Remarkably, we observe a nonmonotonic behavior of $v_L$ with increasing $\left|\epsilon\right|$ in the PW phase, which originates from the roton structure in the excitation spectrum~\cite{JiS_PRL_114_105301,Martone_PRA_86_063621}; see the solid line in figure~\ref{excitation_spectrum}.
In the small $\left|\epsilon\right|$ regime of the PW phase, $v_L$ is still equal to the smaller sound velocity of $c_{1,2}$.
However, due to the appearance of the roton structure, in the larger $\left|\epsilon\right|$ regime $v_L$ is no longer equals with the sound velocity but decided by the roton minimum.
Since the energy of the roton minimum decrease with an increasing $\left|\epsilon\right|$, $v_L$ will be suppressed more and more strongly as $\left|\epsilon\right|$ increases.

Generally speaking, the correlation length $\xi$ is defined by the equality between the kinetic energy per particle $\hbar^2/(2m\xi^2)$ and the interaction energy per particle $g\bar{n}$. However, the Landau critical velocity $v_L$ provides another general definition of the correlation length $\xi$ according to $\xi=\hbar/(mv_L)$, which is consistent with the usual definition~\cite{Giorgini_RMP_80_1215}. Therefore, the Landau critical velocity $v_L$ should have a power-law scaling behavior around the critical point as:
\begin{equation}
v_L\propto\xi^{-1}\propto\left|\epsilon\right|^{\nu}.
\end{equation}
In figure~\ref{Scaling_vl}b, we plot the $v_L$ for different $\left|\epsilon\right|$ near the critical point in a log-log coordinate. It shows clearly that $v_L$ has a power-law dependence on $\left|\epsilon\right|$, which can be expressed by $v_L\propto\left|\epsilon\right|^b$. Through linear fitting, we find $b=0.4956$ and $0.4913$ for the PW and ZM phases respectively. This indicates that the static correlation length critical exponent $\nu=1/2$.

\section{Time-evolution dynamics across the critical point}

In this section, we show how to obtain the Kibble-Zurek scalings from the real-time dynamics.
We perform numerical simulations of spontaneous magnetization and domain formation based on the GPEs~(\ref{GPEs}).
Starting with the ZM phase, we linearly change the Raman coupling strength $\Omega$ to drive the system across the critical point between ZM and PW phases according to
\begin{equation}
\Omega(t)= (1-t/\tau_Q)\Omega_C.
\end{equation}
In our simulation, we adopt various quench times $\tau_Q$ over two orders of magnitude and we perform 100 runs of simulations for each $\tau_Q$. On the other hand, since the quantum fluctuations that trigger the growth of magnetization are ignored in the MF approximation~\cite{Saito_PRA_76_043613}, we introduce appropriate noise to the initial state so that the dynamics of spontaneous magnetization can be studied by the MF theory.
\begin{figure}[htb]
\begin{center}
\includegraphics[width=10 cm]{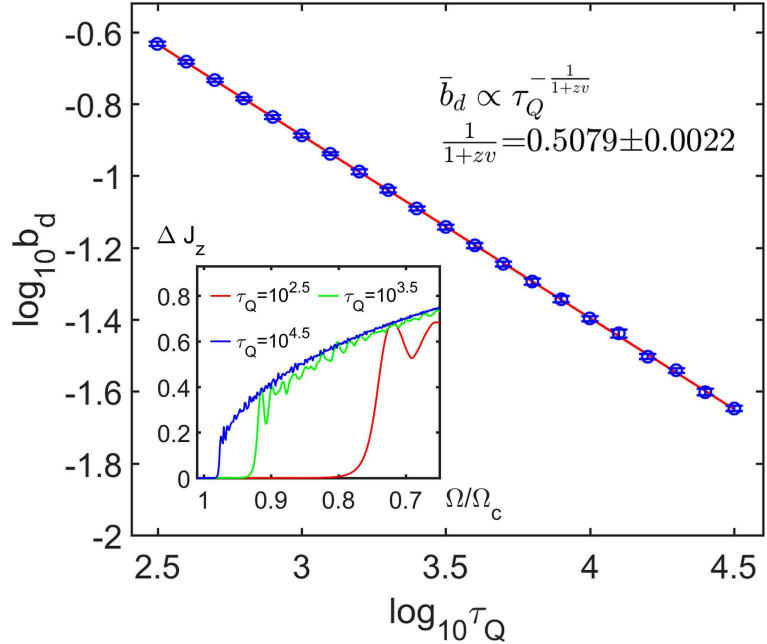}
\caption{\label{bd_scaling} (Color online) The scaling of average bifurcation delay $\bar{b}_d$ with respect to quench time $\tau_Q$. The inset shows the growth of spin fluctuation $\Delta J_z$ with three typical quench times $\tau_Q$. In each run, the system is judged to be unfrozen when $\Delta J_z$ exceeds a threshold $0.05$ and then the bifurcation delay $b_d=\left|\Omega(\hat{t})-\Omega_C\right|$ is obtained. The error bars corresponds to standard deviation of 100 runs. The parameters: $gn/E_r=1$, $\gamma=(g-g_{12})/(g+g_{12})=0.0012$, $L=200$, $N=10^5$. }
\end{center}
\end{figure}

After the Raman coupling strength $\Omega$ sweeping through the bifurcation point of the quantum phase transition, the BECs manifest delayed development of spin fluctuation.
To determine the freeze-out time $\hat{t}$ in each single run, one can utilize fluctuations of the spin polarization
\begin{equation}
\Delta J_z=\sqrt{\frac{1}{L}\int\!J_z^2(x)dx-\left[\frac{1}{L}\int\!J_z(x)dx\right]^2}.
\end{equation}
Since the critical exponents $z$ and $v$ are insensitive to the choice of the thresholds~\cite{Damski_PRL_99_130402}, we adopt the threshhold $\Delta J_z=0.05$ in our numerical results.
We have also checked that the same conclusions can be obtained for other thresholds between $0.01$ to $0.1$.
In figure~\ref{bd_scaling}, we show the bifurcation delay $b_d=\left|\Omega(\hat{t})-\Omega_C\right|$ for different quench time $\tau_Q$.
It is clear that the growth of spin fluctuations lags the phase transition point by an amount of $\Omega$ and the system stays frozen for a larger bifurcation delay $b_d$ for smaller quench time; see the inset of figure~\ref{bd_scaling}.
Such bifurcation delay has also been reported in laser pumped BECs~\cite{LeeC_PRA_69_033611}.
In figure~\ref{bd_scaling}, it is illustrated that the average bifurcation delay $\bar{b}_d$ fits well to a power-law scaling with respect to the quench time $\tau_Q$, which yields an exponent $-0.5079$.
Since $b_d\propto\hat{\epsilon}$, we obtain $1/(1+z\nu)=0.5079$ according to (\ref{scaling_1}).

\begin{figure}[htb]
\begin{center}
\includegraphics[width=10 cm]{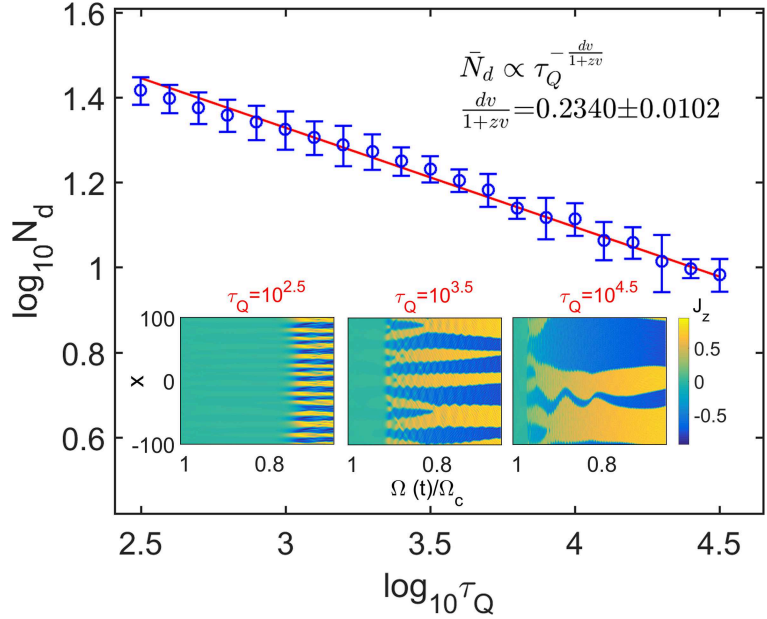}
\caption{\label{Nd_scaling} (Color online) The scaling of average domain number $\bar{N}_d$ with respect to quench time $\tau_Q$. The insets show the growth of domains with three typical quench times $\tau_Q$. In each run, we count the domain number $N_d$ at the freeze-out time $\hat{t}$, which is defined as the time when $\Delta J_z$ exceeds a threshold $0.05$. The error bars corresponds to standard deviation of 100 runs. The parameters are the same with those for figure~\ref{bd_scaling}. }
\end{center}
\end{figure}

In order to extract the critical exponents, we also analysis the universal scaling of domain number $N_d$ versus quench time $\tau_Q$.
Typical examples of domain-formation dynamics for different quench times are illustrated in the insets of figure~\ref{Nd_scaling}.
One can see that ferromagnetic domains form after the system crossing the phase transition point and the average domain size increase with the quench time.
In each run, we count the domain number $N_d$ by identifying the number of zero crossings of $J_z(x)$ at the unfreezing time $\hat{t}$.
The domain numbers for different $\tau_Q$ are summarized in figure~\ref{Nd_scaling}.
We observe that the average domain number follows a power-law scaling $\bar{N}_d\propto\tau_Q^{-0.2340}$ as expected from the KZM, which gives the scaling exponent $d\nu/(1+z\nu)=0.2340$.
We have checked that similar scaling of the average domain numbers can be obtained for other thresholds of $\Delta J_z$ between $0.01$ to $0.1$.

Finally, combining the scaling exponents of the average bifurcation delay and the average domain number with respect to $\tau_Q$, we obtain the critical exponents $\nu=0.4607$ and $z=2.1030$, which consist with the analytical exponents obtained in Sec. \uppercase\expandafter{\romannumeral3} and agree well with the MF exponents $\nu=1/2$ and $z=2$.

\section{Conclusions and discussions}

In summary, we have investigated the universal spatiotemporal dynamics across a second-order phase transition point in SO coupled two-component BECs.
Due to the divergence of correlation length and relaxation time at the critical point, spatial domains form in the phase transition dynamics according to the KZM.
We analyze the Bogoliubov excitation spectrum and find the Landau critical velocity vanishes at the phase transition point, which results in the spontaneous appearance of the elementary excitations.
We extract the critical exponents from the excitation spectrum and the scaling of the Landau critical velocity around the critical point.
On the other hand, we numerically find that the average bifurcation delay and the average domain number after the system crossing the critical point follow a universal scaling law as expected by the KZM.
We also extract the critical exponents from the numerical scalings.
The critical exponents given by the two methods consist with each other.

Based upon current available techniques for SO coupled BECs, it is possible to probing the above KZ scalings.
The SO coupling can be synthesized in two-component BECs with two counter propagating Raman lasers~\cite{LinY_Nature_471_83}.
The plane-wave phase and the zero-momentum phases have been observed in present experiments~\cite{JiS_NP_10_314} and the high tunability of the Raman coupling parameters make the quenching across the phase transition point possible.
The KZ scalings can then be obtained by measuring the bifurcation delay and the size of ferromagnetic domains for different quench rates via the time-of-flight~\cite{Clark_Science_354_606}.

\begin{ack}

This work was supported by the National Natural Science Foundation of China (Grants No. 11374375, No. 11574405).

\end{ack}

%\bibliographystyle{unsrt}
%\bibliography{KZM_SOCBEC}

\end{document}